\documentclass[twocolumn,aps,prl,showpacs]{revtex4}
\usepackage{epsfig}  

\newcommand{\gl}[1]{Eq. (\ref{#1})}
\newcommand{\gls}[2]{Eqs. (\ref{#1},\ref{#2})}

\def\gtrless{\raise2.5pt\hbox{$>$}\llap{\lower2.5pt\hbox{$<$}}}
\def\gtrapprox{\raise2.5pt\hbox{$>$}\llap{\lower2.5pt\hbox{$\approx$}}}
\newcommand{\bsq}[1]{\begin{subequations}\label{#1}}
\newcommand{\esq}{\end{subequations}}
\newcommand{\beq}[1]{\begin{equation}\label{#1}}
\newcommand{\eeq}{\end{equation}}
\newcommand{\beqa}[1]{\begin{eqnarray}\label{#1}}
\newcommand{\eeqa}{\end{eqnarray}}

\newcommand{\vek}[1]{{\bf #1}} 
\newcommand{\gd}{\dot{\gamma}}
\newcommand{\kap}{\mbox{\boldmath $\kappa$}}

\newcommand{\smopb}{\Omega_B^{(\gd)}}
\newcommand{\nix}{{\phantom{.}}}
\renewcommand{\rho}{\varrho}

\def\gtrapprox{\raise2.5pt\hbox{$>$}\llap{\lower2.5pt\hbox{$\approx$}}}
\def\lessapprox{\raise2.5pt\hbox{$<$}\llap{\lower2.5pt\hbox{$\approx$}}}

\begin{document}

\title{Theory of nonlinear rheology and  yielding of dense colloidal
suspensions} 
 
\author{Matthias Fuchs$^*$ and Michael E. Cates}

\affiliation{Department of Physics and Astronomy, The University of
Edinburgh, JCMB King's Buildings, Edinburgh EH9 3JZ, GB}
\altaffiliation{Permanent address:
Physik-Department, Technische Universit{\"a}t M{\"u}nchen,   85747
Garching, Germany.}  

\date{\today}


\begin{abstract}
A first principles approach to the nonlinear flow of dense
suspensions is presented which captures shear thinning of colloidal
fluids and  dynamical yielding of colloidal glasses. The advection of
density fluctuations plays a central role, suppressing the caging of
particles and speeding up structural relaxation. 
A mode coupling approach
is developed to explore these effects.
\end{abstract}
\maketitle

The properties of dispersions under flow are central to their processing
and technological use \cite{larson,russel}. But especially the
non-linear rheology  is not yet well understood.
For the simplest
case of steady shearing, the low density behavior is known
\cite{Blawzdziewicz93}, but upon increasing the density
the growing importance 
of particle interactions requires theoretical approximation
\cite{Lionberger00,Bergenholtz01b}, hinders simulations
\cite{Foss00}, and calls for studies of
model systems, e.g. \cite{Laun92,Nommensen99}.
Of major interest is the arrest of the structural relaxation 
when approaching solidification for higher densities, which raises the
question of how the imposition of steady shearing might interfere with
glass formation.  The linear  phenomenology is familiar:
a colloidal fluid possesses a viscosity and flows,  while a
colloidal glass characterized by elastic constants, only distorts under
strain \cite{larson,russel}. But the nonlinear rheology of glassy
colloids, which exhibit a 
continuous slowing down of the structural relaxation due to particle
blocking (the ``cage effect'') \cite{larson}, is less clear.
While the mode coupling theory (MCT) recovers the linear
phenomenology of this fluid-to-glass transition from microscopic
starting points \cite{Bengtzelius84}, a nonlinear
external driving introduces new time scales whose
influence on (non-)equilibration is not understood,
and has been addressed only in minimal models \cite{Fielding00} or
mean-field approaches \cite{Berthier00}.  
Moreover, as the true nature of the glass is still uncertain, its behavior
under shearing may provide broader new insights (as suggested by
recent simulation studies \cite{Yamamoto98,Barrat01}).  

Here we develop a first-principles approach for the simplest case of a
disordered colloidal suspension under steady imposed shear, neglecting both
many-body hydrodynamics and the resulting velocity fluctuations. We
first identify some generic features in the yield properties of glass;
approximations suggested by the MCT are then introduced in order to
derive quantitative predictions.  

The system consists of $N$ spherical
particles (diameter $d$) 
dispersed in a volume $V$ of solvent with imposed flow
profile $\vek{v}(\vek{r}) = \kap\,  \vek{r}$,  where for 
simple shear with velocity along the $x$-axis and
its gradient along the $y$-axis, the shear rate tensor is
$\kappa_{ij}=\gd\  \delta_{ix}\delta_{jy}$. The effect of the shear
rate $\gd$ on the particle dynamics is measured by the Peclet number
\cite{russel}, Pe$_0=\gd d^2/D_0$, formed with the bare diffusion
coefficient $D_0$ of a particle.  
Dimensionless units are obtained by setting
$d=D_0=k_BT=1$.
The evolution of the distribution function $\Psi$ of the particle positions, 
$\vek{r}_i$, $i=1,\ldots N$, under internal forces $\vek{F}_i$ and
shearing, but neglecting hydrodynamic interactions, is given by the
Smoluchowski equation \cite{russel,dhont}: 
\beq{e1}
\partial_t \Psi = \Omega^{(\gd)} \Psi\; \mbox{, where }\;
\Omega^{(\gd)} = \sum_i \vek{\partial}_i \cdot ( \vek{\partial}_i -
\vek{F}_i -  \kap \,\vek{r}_i ) \; .
\eeq

The system is taken to be in quiescent equilibrium ($\gd=0$) at $t\le0$
when averages $\langle \ldots \rangle^{(\gd=0)}$ are the canonical
equilibrium ones. Then at $t=0^+$, the velocity profile is 
switched on instantaneously, so that the steady state distribution
function $\Psi_s$, which satisfies $\Omega^{(\gd)}
\Psi_s = 0$, will be approached at long times, $t\to\infty$.
If $\Psi_s$ was known the steady state average $\sigma=\langle
\sigma_{xy} \rangle^{(\gd)}$ of the 
(thermodynamic) shear
stress \cite{Batchelor77} could be found. From this the shear
viscosity would follow as $\eta(\gd) = \eta_\infty+\sigma/\gd$ (where
the solvent viscosity is denoted $\eta_\infty$). But the  
rapid increase of $\eta(0)$ close to the glass transition
suggests that $\Psi_s$ is sensitively dependent on shear rate,
which makes a direct calculation of it difficult. 

A more robust way to approximate steady-state quantities comes from
the insight that the growth of $\eta$ at  
the glass transition arises by slowing down of structural
relaxations \cite{Bengtzelius84},
 whose characteristic time $\tau$ in the quiescent
state defines a second, ``dressed'' Peclet (or Weissenberg
\cite{larson}) number, 
Pe $=\gd\tau$. This characterizes the influence of shear on structural
relaxation and increases without bound at the glass transition, even
while Pe$_0\ll 1$. We argue that the competition of structural
rearrangement and shearing that arises when Pe$> 1 \gg $Pe$_0$ dominates the
non-linear rheology of colloids near the glass transition. Therefore
steady state quantities shall be determined by considering the
structural relaxation under shearing and ``integrating through the 
transient dynamics''.
Because Pe$_0\ll1$, we expect ordering or layering transitions to be
absent \cite{Laun92}; and as hydrodynamic interactions are presumed to
play a subordinate role during the structural relaxation
\cite{Bergenholtz01b} we neglect these too, focusing solely on the
Brownian contribution to the transverse (shear) stress. 

With $t_0$ the time passed since the start of shearing, the
correlation function of fluctuations in variables $f$ and $g$
separated by time $t$ is given by:
\beq{e2}
C_{fg}(t,t_0) 
   = \langle
 e^{\Omega_B^{(\gd)}\, t_0 }\,  f^*\; 
\left( e^{\Omega_B^{(\gd)}\, t }\, g \right) \rangle^{(0)}\; ,
\eeq
where the backwards Smoluchowski operator $\smopb$ arises from
 partial integrations:
$\smopb = \sum_i ( \vek{\partial}_i +
\vek{F}_i + \vek{r}_i \kap^T)  \cdot  \vek{\partial}_i$, and $t,t_0>0$.
Steady state expectation values,
$\langle g\rangle^{(\gd)}=C_{1g}(t,t_0\to\infty)$, and variances,
$\langle f^* g\rangle^{(\gd)} = C_{fg}(t=0,t_0\to\infty)$ then follow.

\begin{centering}
\begin{figure}[h]
\centerline{\psfig{file=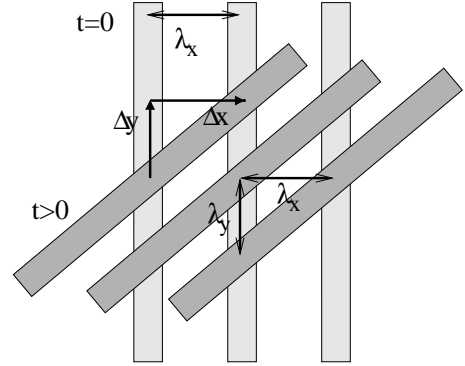,width=6.cm}}  
\caption{Advection by steady shear of a
fluctuation in $x$-direction  with wavelength $\lambda_x$ at $t=0$.
At later time $t$, its wavelength $\lambda_y$ in $y$-direction obeys:
$\lambda_x/\lambda_y=\Delta x / \Delta y= \gd t$.} 
\label{fig1}
\end{figure}
\end{centering}

An important property of the sheared system is 
translational invariance \cite{Onuki79}. It leads to
spatially independent averages, or, in Fourier space at
wavevector $\vek{q}$, to
$\langle f_\vek{q}\rangle^{(\gd)}=\langle f_0 \rangle^{(\gd)}\,
\delta_{\vek{q},0}$.
In the two-time correlation functions of \gl{e2}, it leads to a
coupling of fluctuations of wavevector $\vek{q}$ with later fluctuations
of the {\em advected wavevector} $\vek{q}(t)=
\vek{q}+\vek{q}\kap \,t$, suggesting  the definition: 
$C_{f_\vek{k} g_\vek{q}}^\nix(t,t_0) = N C_{fg;\vek{q}}(t,t_0) \,
\delta_{\vek{q}(t),\vek{k}}$.
Figure \ref{fig1} sketches the advection of a fluctuation with
initial wavelength $\lambda_x$ to one with $\lambda_x$ and
$\lambda_y(t)=\lambda_x/(\gd t)$ at later time $t$. Brownian particle
motions (assisted by the interaction forces) ``smear out'' the
fluctuation with time and cause the decay of the corresponding
correlator. Because the wavenumber $q(t)$ increases upon shearing,
smaller and smaller motions can cause the 
fluctuation to decay \cite{Onuki79}.

Equation (\ref{e2}) is an exact consequence of \gl{e1}
(for the given shear flow); but to
proceed further requires additional
approximations. With our assumption that
applied shear interacts with slow structural rearrangements, we build
on the description achieved by the MCT, and analyse the approach into
the steady state by monitoring the fluctuations of 
density ($\rho_\vek{q}=\sum_{i=1}^N e^{i\vek{q}\vek{r}_i}$)
and of the ``pair density'' (the square of the density in real space),
aiming to establish nonlinear closed equations for  
these. This entails elimination of forces $\vek{F}_i$ in favor of the
quiescent-state structure factor $S_q$ (taken to be known) --  a
near-equilibrium assumption (see below) that is formally uncontrolled
but motivated, at least in part, by the smallness of Pe$_0$.   

Steady state correlators are now 
approximated by projection onto the density modes, giving
\beq{e4}
\langle f^* g\rangle^{(\gd)}  \approx 
\langle f^* g  \rangle^{(0)} + 
\frac{\gd}{2}  \int_0^\infty\!\!\!\!\!\!dt\, 
\sum_{\vek{k}}
\frac{k_xk_y(t)S'_{k(t)}}{k(t)NS^2_{k(t)}}\; V^{fg}_{\vek{k}} \;
\Phi^2_{\vek{k}}(t)\; ,
\eeq
with $t$ now the time since switch-on, $S'_k=\partial S_k/\partial k$, and 
$V^{fg}_{\vek{k}}$ an easily found static
overlap  \cite{overlap}.
The transient density fluctuations are given by
$\Phi_{\vek{q}}(t)=C_{\varrho\varrho;\vek{q}}(t,0)/S_q$, and are
normalized by $S_q$. 
They enter \gl{e4} via a factorization--approximation of the density
pair fluctuation functions.
With the choice $f=g=\rho_\vek{q}/\sqrt N$, \gl{e4} gives the steady
state structure factor under shear, whereas  
$f=1$ and $g=\sigma_{xy}/V$ give the transverse stress, for which
$V^{fg}_{\vek{k}} = (Nk_xk_y/Vk)  S'_k$.

The problem of calculating the steady state  averages is thus ``reduced''
to first finding the transient density fluctuations
$\Phi_{\vek{q}}(t)$,  given by the structural
rearrangements  after switching on the flow, and integrating
these in  \gl{e4}. From Zwanzig-Mori (type)
manipulations \cite{Kawasaki95}, one finds the exact equation of motion:
\beq{e5}
\dot{\Phi}_\vek{q}(t) + \Gamma_\vek{q}(t)  \left\{
\Phi_\vek{q}(t) + 
 \int_0^t\!\!\! dt'\; m_\vek{q}(t,t') \,
\dot{\Phi}_\vek{q}(t')
\right\} = 0 \; .
\eeq
where $\dot{\Phi}_\vek{q}(t)=\partial_t \Phi_\vek{q}(t)$, and
the ``initial decay rate'' $\Gamma_{\vek{q}}(t)$ exhibits the
familiar Taylor dispersion \cite{dhont,taylorconv}. It is not known
how to evaluate the microscopic expression for the memory
function $m_\vek{q}(t,t')$ exactly. In the MCT spirit of our
approach, this is approximated by projecting the fluctuating forces onto
density pairs and factorizing the resulting pair-density correlation
functions as
\beq{e6}
 m_\vek{q}(t,t') \approx 
\frac{1}{2N} \sum_{\vek{k}}  V^{(\gd)}_{\vek{q},\vek{k}}(t,t')\;
 \Phi_{\vek{k}}(t-t') \, \Phi_{\vek{q}-\vek{k}}(t-t') \; .
\eeq
The vertex $ V^{(\gd)}$, whose lengthy formula will be published elsewhere,
is evaluated in the limit Pe$_0\ll1$ (as argued above) but for
large times so that 
$\gd t$ and $\gd t'$ are  finite.
As $\gd\to 0$, it reduces to the standard MCT vertex
\cite{Bengtzelius84} and like the latter is uniquely determined by the
equilibrium structure factor, $S_q$. For long times, it vanishes
as $V^{(\gd)}\propto q_y q_x^{-3} \gd^{-3} {t'}^{-2}(t-t')^{-1}$ for $q_x\ne0$.

Equations (\ref{e4}) to (\ref{e6}) complete our derivation 
of  closed, self-consistent equations for the steady state
properties of dense sheared suspensions. They contain the  bifurcation
singularities which lie 
at the core of MCT. For $\gd=0$, upon smooth changes of the input equilibrium
state parameters, a fluid 
with $\Phi_q(t\to\infty)\to0$ turns into an amorphous solid,
$\Phi_q(t\to\infty)\to f_q>0$. The $f_q$ are called glass form factors
and describe the arrested structure. While transport
coefficients of the fluid, like the viscosity, are connected to the
longest relaxation  time of $\Phi_q(t)$, elastic constants of the
solid, like the  transverse elastic modulus $G_\infty$, are given by the
$f_q$ \cite{Bengtzelius84}.

In the limit of  small shear rates, a stability analysis of the
amorphous solid can be performed and leads to a generalization of the
factorization theorem of MCT \cite{Goetze85}. Close to the
bifurcation, the dynamics are governed by 
$ \Phi_q(t) = f^c_q + h_q \,   {\cal G}(t)$,
where the $f_q^c$ describe the glassy structure at the instability and
the critical amplitude $h_q$ is connected to the cage-breaking
particle motion; both retain their definition from the unsheared situation.  
Here ${\cal G}(t)$ contains the essential
non-linearities of the bifurcation dynamics which arise from the physical
feedback mechanism (the cage effect). It
depends on a few material parameters only and, for $|{\cal G}(t)|\ll
1$, follows from: 
\beq{e7}
\varepsilon - c^{(\gd)} \; (\gd\; t)^2 + \lambda\; {\cal G}^2(t) =
\frac{d}{dt} \; \int_0^tdt'\; {\cal G}(t-t')\; {\cal G}(t') \; .
\eeq
Here, $\varepsilon$ measures the distance to the transition 
 and  $\lambda$ is known for some systems \cite{Bengtzelius84}. Monte Carlo
estimations of the microscopic expression for
 the (new) parameter $c^{(\gd)}$ give
$c^{(\gd)}\approx3$ for hard spheres.
Corrections of higher order in the small quantities ($\varepsilon,
\gd, {\cal G}$) are
neglected; see \cite{Franosch97} for background on \gl{e7} for $\gd=0$.

Equation (\ref{e7}) is our central result.
As expected, the sign of $\gd$ does not enter, although it
affects the Taylor dispersion.
Because $(\gd t)^2$ dominates for long times, always
${\cal G}(t\!\!\to\!\!\infty)\to-t/\tau^{(\gd)}$, with
$\tau^{(\gd)}=\sqrt{(\lambda-1/2)/c^{(\gd)}}/  |\gd| $.  Hence, under
flow, density fluctuations always decay, as this decrease of ${\cal G}(t)$ for
long times initiates the final relaxation  (where the corrections to
\gl{e7} become important) of  $\Phi_{\vek{q}}(t)$ to zero.
Arbitrarily small shear rates $\gd$ melt the glass and so ``interrupt'' aging,
as has also been found for spin-glasses where
shearing was mimicked by breaking detailed balance \cite{Berthier00}. 
This vindicates our decision to ignore aging and to proceed via  
\gl{e4} in order to obtain steady state properties.

While the non-Newtonian fluid behavior ($\varepsilon<0$) includes two
slow time scales, the familiar $\tau$ and the shear induced $\tau^{(\gd)}$, 
the rheology of glass ($\varepsilon\ge0$) is determined by
$\tau^{(\gd)}$ only \cite{Fuchs02}.
For $\varepsilon\to0+$ and $\gd = 0$, a fraction $f_q=f_q^c+
h_q\,\sqrt{\varepsilon/(1-\lambda)}$ of the density fluctuations
would stay arrested, while with shear these decay at a rate set by $\gd$:
$\Phi_{\vek{q}}(t)\to f_q\, \Phi^{+}_{\vek{q}}(t/\tau^{(\gd)})$,
where $\Phi^{+}_{\vek{q}}(x\!\!\to\!\!0)-1\propto -x$, as follows
from \gl{e7}. 

For $\varepsilon\ge0$ and $\gd\to0$, the time for the final decay
can become arbitrarily slow  compared to the time
characterizing  the decay onto $f_q$. 
Inserting $\Phi^{+}_{\vek{q}}(t/\tau^{(\gd)})$ into
\gl{e4}, the long time contributions separate out, and depend on time
only via $\gd t$. Hence the glass has nontrivial
shear-rate-independent limits for steady 
state values: $\langle f^* g\rangle^{(\gd)}\to \langle f^*
g\rangle^{(+)}  \ne \langle f^* g\rangle^{(0)}$ for
$\gd\to0$. The $ \langle f^* g\rangle^{(+)}$ are given by integrals over the
$\Phi^{+}_{\vek{q}}(t/\tau^{(\gd)})$ and quantify those fluctuations
that require 
the presence of shearing to avoid their arrest.
For the case of the shear stress, a finite
(dynamical) yield stress, $\sigma^+=\lim_{\gd\to0} \langle
\sigma_{xy}\rangle^{(\gd)}$ 
for $\varepsilon\ge0$, is thereby found. Since the glass transition is
often identified by a divergence of viscosity (in terms of which we
have shear thinning: $\eta(\gd\to0) \propto \gd^{-m}$ with $m = 1$)
our prediction of a finite yield stress throughout the glass phase is
far from trivial. It excludes e.g. power-law fluid behavior
($\sigma\sim\gd^{1-m}$ with $0<m<1$; see \cite{Fielding00}).  

These results follow from the general stability analysis of
\gl{e7} and are predicted to be universal, i.e. to hold 
for the Brownian contribution to the shear stress and
viscosity close to the glass transition in any colloidal dispersion.
But because (in contrast to aging approaches \cite{Latz02}) we
approximate nonlinear couplings under shear using 
equilibrium averages, we require the system to remain ``close to
equilibrium'' in some sense. The existence of a finite yield stress
$\sigma^+$ means that this is not guaranteed even as $\gd \to 0$.

Insight into an important mechanism of shear-fluidization can be
gained by considering transient density fluctuations with   
wavevector $\vek q$  perpendicular to the flow plane, $\vek{q}=q\hat{e}_z$.
Here $V^{(\gd)}$ simplifies to the standard MCT vertex with advected
(time-dependent) wavevectors \cite{isovertex}. While  for
$\gd=0$ it exhibits the nonlinear coupling of density
correlators with wavelength equal to the average particle distance,
for $\gd\ne0$, the (only)  effect of shearing on this ``neutral'' direction
consists in a shift of the advected wavevectors 
to higher values, where the effective potential decreases. This
decreases the memory function and thus speeds up structural rearrangements.
In this way the theory captures the faster decay
of fluctuations caused by shear advection (cf Fig. \ref{fig1}).  

\begin{centering}
\begin{figure}[h]
\centerline{\psfig{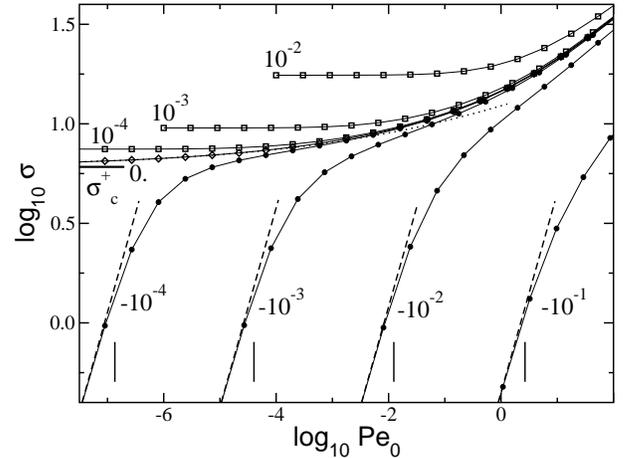}}  
\caption{Steady shear stress,
 $\sigma=\langle\sigma_{xy}\rangle^{(\gd)}$,  in units of $k_BT/d^3$ 
 versus Pe$_0=\gd d^2/D_0$, for a model of hard spheres
\protect\cite{isomodel} at various distances from its glass
transition, $(\phi-\phi_c)$ as labeled. For the fluid cases,
$\phi<\phi_c$,  
dashed lines indicate Newtonian fluid behavior, $\sigma=\eta\gd$,
while vertical bars mark Pe$=\gd\tau=1$, with the structural relaxation
time taken from $\Phi_{q=7/d}(t=\tau)=0.1$. For the
critical density, $\phi_c$, the critical yield stress, $\sigma^+_c=6.0$,
is shown by a horizontal bar, and the dotted line
$\sigma=\sigma^+_c(1+1.0 \gd^{0.17})$ matches for $\gd\to0$.}
\label{fig2}
\end{figure}
\end{centering}

The presence of shear advection in  the neutral $z$-direction
suggests an approximation that considers only the resulting competition
of caging and advection-induced decay.  In this ``isotropically sheared
hard sphere model'' (ISHSM) \cite{isomodel}, we neglect kinematic flow
of particles so that all directions are treated as 
``neutral''.  The quiescent $S_q$ depends only on the packing
fraction $\phi$, and
 the model's glass
transition lies at $\phi_c=0.51591$ \cite{Franosch97}.
Figure \ref{fig2} shows the
stress versus strain rate curves
for $\phi$ close to the transition. In the fluid, $\phi<\phi_c$, a
Newtonian regime ($\sigma=\eta \gd$) is found for $\gd$ small
enough that Pe $<1$.
For Pe $>1$
there is a broad crossover to the critical yield stress
value, $\sigma^+_c= \sigma^+(\varepsilon\to0+)$, from which $\sigma$
starts to rise  due to (non-universal) short-time effects for Pe$_0$
around $10^{-2}$ (where 
$\eta_\infty$ will also contribute).  In the glass,
$\phi\ge\phi_c$, a yield stress plateau for $\gd\to0$ is obtained, 
and rises strongly with increasing packing fraction.  
We speculate that the lack of a clear yield stress
plateau at Pe$_0\ge10^{-3}$ (and likely hydrodynamic effects) explains
shear-thinning exponents $m<1$ seen in experiments \cite{Laun92}.

In summary,  we have presented a microscopic theory of the
nonlinear rheology of colloidal fluids and glasses under steady
shear. It predicts a universal transition between shear-thinning
fluid flow, with diverging viscosity upon increasing the interactions,
and solid yielding, with a yield stress that is finite at (and beyond)
the glass point. 
Besides its interest for dispersion flow, our work suggests
a further role of colloidal systems in elucidating
glasses via the study of shear-melted states for small shear rates. 
This is of fundamental interest
because, e.g., mean-field driven spin-glass theories predict nothing
like a yield stress \cite{Berthier00}. While  
comparison with measurements in colloids \cite{Foss00,Laun92,Nommensen99} and
simulations of sheared atomic glasses
\cite{Yamamoto98,Barrat01} is
promising, our approach represents only the first step toward rational
prediction of the rheology of a glass;  though physically motivated
and in part inspired by the successful MCT description of the cage-effect,
 several of our approximations remain incompletely justified. 
Nor is it clear whether \gls{e5}{e6}
can exhibit ``jamming'' transitions \cite{Cates98} at finite shear
rates, or how 
anisotropic \cite{Cates89} the fluctuations can become. Extension to
time dependent shearing would be especially interesting 
because recent shear echo measurements \cite{Petekidis02} reveal
intriguing glass-melting scenarios.
 
\begin{acknowledgments}
We thank J.-L. Barrat, J. Bergenholtz, L. Berthier,
A. Latz and G. Petekidis for discussions. 
M.F.\ was supported by the DFG, grant Fu~309/3. 
\end{acknowledgments}


\begin{thebibliography}{10}
\bibitem{isomodel}
The ISHSM replaces $k_z$ by $\vek{q}\!\!\cdot\!\!\vek{k}/q$ in 
$V^{(\gd)}_{q\hat{e}_z,\vek{k}}(t,t')$, and uses this vertex
for all $\vek{q}$ with the approximation
$k(t)=k(1+(t\gd)^2/3)^{1/2}$; also
$\Gamma_{\vek{q}}(t)=q^2/S_q$. The $\Phi_{\vek{q}}(t)$ become
isotropic, and Eqn. (\protect\ref{e4}) for the stress is approximated by
  $\sigma=(\gd/(60\pi^2))\int_0^\infty\!\!dt  \int\!\! dk\, k^4\Phi_k^2(t)
  S'_kS'_{k(t)}/S^2_{k(t)}$.  Then the model is handled as in
  \protect\cite{Franosch97}, giving $\lambda=0.74$ and
$\varepsilon= 1.54 (\phi-\phi_c)/\phi_c$. It
underestimates shearing, as the
ratio  $\sigma^+_c/G^c_\infty=0.33$, is overestimated
\protect\cite{russel,larson,Nommensen99}.
\end{thebibliography}
\end{document}